\documentclass[showpacs,twocolumn,aps,prl]{revtex4-2}
\usepackage{graphicx}
\usepackage{dcolumn}
\usepackage{bm}
\usepackage{amsmath}
\usepackage{amssymb}
\usepackage[usenames,dvipsnames]{xcolor}
\usepackage{color}

\begin{document}
\title{Fundamental Distinction between Intrinsic and Extrinsic Nonlinear Thermal Hall Effects}
\author{Da-Kun Zhou$^{1}$}

\author{Zhi-Fan Zhang$^{1}$}

\author{Xiao-Qin Yu$^{2}$}
\email{yuxiaoqin@hnu.edu.cn}

\author{Zhen-Gang Zhu$^{1,3,4}$}
\email{zgzhu@ucas.ac.cn}

\author{Gang Su$^{1,4,5}$}
\email{gsu@ucas.ac.cn}

\affiliation{$^{1}$ Theoretical Condensed Matter Physics and Computational Materials Physics Laboratory, College of Physical Sciences, University of Chinese Academy of Sciences, Beijing 100049, China.\\
$^{2}$School of Physics and Electronics, Hunan University, Changsha 410082, China.\\
$^{3}$School of Electronic, Electrical and Communication Engineering, University of Chinese Academy of Sciences, Beijing 100049, China. \\
$^{4}$ CAS Center for Excellence in Topological Quantum Computation, University of Chinese Academy of Sciences, Beijing 100190, China.\\
$^{5}$ Kavli Institute of Theoretical Sciences, University of Chinese Academy of Sciences, Beijing 100049, China.}

\begin{abstract}
We theoretically investigated the fundamental distinction between intrinsic and extrinsic nonlinear thermal Hall effect in the presence of disorder at the second-order response to the temperature gradient in terms of the semi-classical Boltzmann equation. We found that, at low temperatures, the intrinsic contribution of the nonlinear thermal Hall conductivity is proportional to the square of temperature, whereas the extrinsic contributions (side-jump and skew-scattering) are independent of temperature. This distinct dependency on temperature provide a new approach to readily distinguish the intrinsic and extrinsic contributions. Specifically, we analysed the nonlinear thermal Hall effect for a tilted two-dimensional massive Dirac material. In particular, we showed that when the Fermi energy is located at the Dirac point, the signal is solely from the intrinsic mechanism; when the Fermi energy is higher, the extrinsic contributions are dominant, which are uncovered to be two to three orders of magnitude larger than the intrinsic contribution.
\end{abstract}


\maketitle
Traditionally, Hall (Nernst) effects are studied in linear-response regime, namely, the generated transverse voltage is linearly proportional to driving forces (electric field or temperature gradient). Broken time reversal symmetry due to the presence of magnetization or external magnetic field is required to guarantee such a Hall voltage. 
Recently, a nonlinear anomalous Hall effect \cite{Sodemann} (NAHE) as a second-harmonic response to an ac electric field was proposed and has attracted broad interests in the study of nonlinear anomalous transport phenomena in time-reversal invariant but inversion symmetry broken materials. The most interesting point is that this NAHE is due to the emergent Berry curvature dipole in the momentum space rather than Berry curvature itself. This reveals the complicate interplay between nontrivial topological structure of the energy bands and the transverse Hall-like transport \cite{Sodemann,Z. Z. Du,Low,Raffaele,Facio,You}. NAHE has been predicted in a lot of noncentrosymmetric materials, such as bilayer WTe$_{2}$ \cite{Z. Z. Du}, strained graphene \cite{Raffaele}, and topological crystalline insulator SnTe \cite{Lau}, etc, and has been successfully observed in bilayer \cite{Q. Ma} and few-layer \cite{Kang} WTe$_{2}$.

Although, the origin of anomalous Hall effect has been thought to be from intrinsic \cite{Karplus} (disorder-free) or extrinsic (disorder-induced) contribution which includes side-jump \cite{Berger} and skew-scattering \cite{Smit}, it is still a challenge to identify these contributions experimentally. So far, there are two main approaches, namely the traditional \cite{Karplus,Berger,Smit,Jellinghaus, Dheer, Lee, Y. Pu, A. Fert, C. G. Zeng, Kotzler, Lavine} and the new scaling law \cite{Y. Tian, D. Hou, D. Yue}, to distinguish different mechanisms. In the conventional scaling law, the skew-scattering contribution to anomalous Hall effect can be easily distinguished from the intrinsic and side-jump contributions
through the scaling relations between the induced Hall resistance and the longitudinal resistivity $\rho_{xx}$, namely
$\rho_{int} \varpropto \rho_{xx}^{2}$ \cite{Karplus}, $\rho_{sj} \varpropto \rho_{xx}^{2}$ \cite{Berger}, $\rho_{sk} \varpropto \rho_{xx}$ \cite{Smit}.  However, the contribution from intrinsic mechanism or side-jump cannot be further identified.
Unlike the traditional scaling law in which the scaling $\rho_\text{AH}=f(\rho_{xx})$ depends only on the single-scatter-induced $\rho_{xx}$, the new scaling $\rho_\text{AH}=f(\rho_{i},\rho_{xx})$ (where $\rho_{xx}=\sum_{i}\rho_{i}$) considers the involved multiple competing scatterings and also depends on the partial longitudinal resistivity $\rho_{i}$ stemming from the involved scattering sources \cite{Y. Tian, D. Hou, D. Yue}, such as phonons or impurities.
In experiments, the partial resistivity generated by different scattering sources needs firstly be determined and , then, to determine the contribution of each mechanism through fitting scaling parameters.
Recently, Lu \textit{et al.,} applied the basic idea of the new scaling law to the nonlinear response regime \cite{Du} to distinguish the intrinsic and extrinsic contribution to the nonlinear anomalous Hall effect.

\begin{figure}[htbp]
\centering
\includegraphics[width=0.44\textwidth]{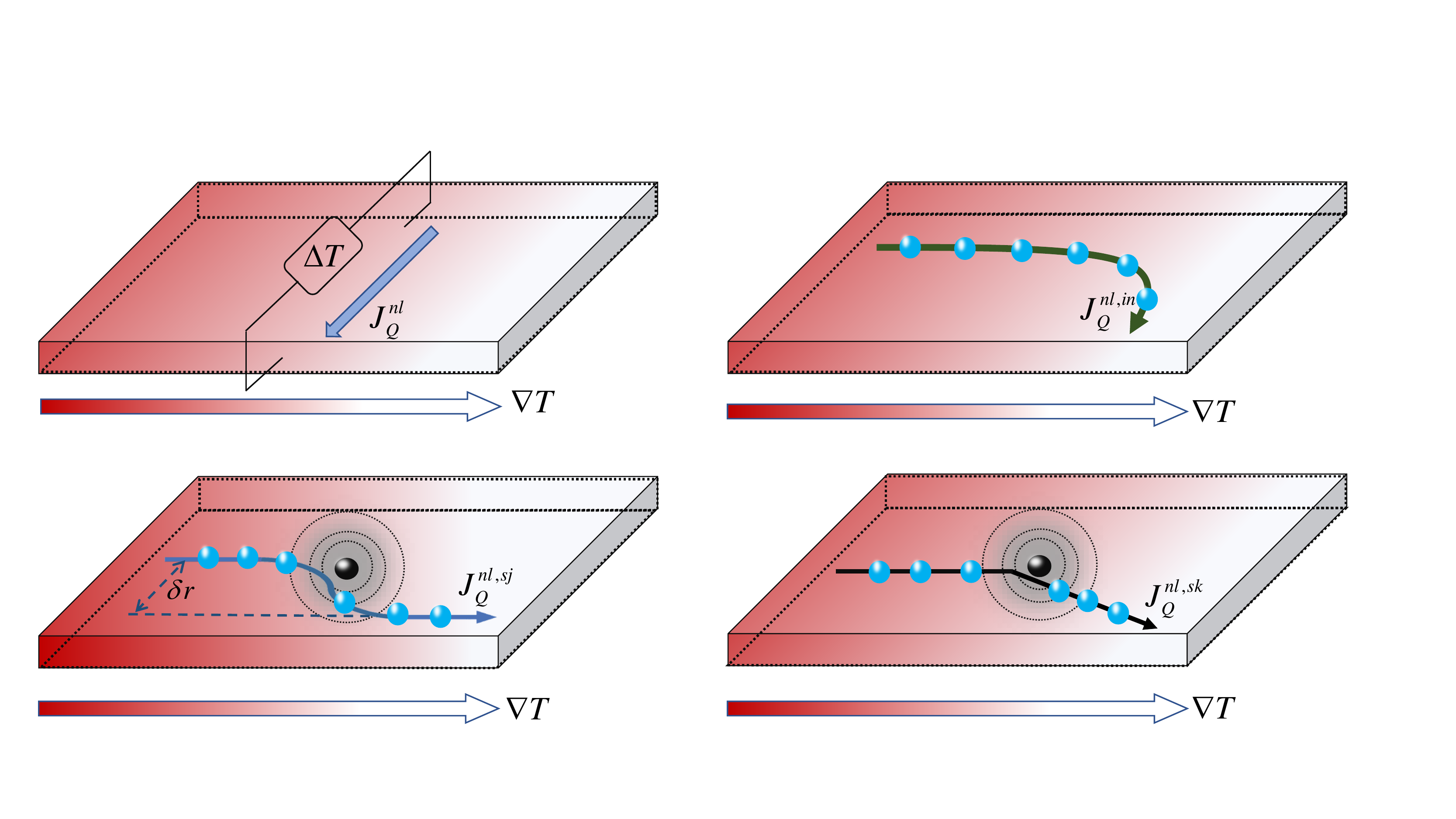}
\centering
\caption{(a) Schematic illustration of the nonlinear thermal Hall effect (NTHE) as a second-order response to the temperature gradient, the nonlinear thermal Hall current (NTHC) $\mathbf{j}^{nl}\propto (\nabla T)^{2}$. The generation of NTHE stemmed from intrinsic contribution (b), side-jump (c) and skew-scattering (d). The black ball in (c) and (d) represent disorder.}
\label{fig1}
\end{figure}

The thermal Hall effect refers to a generation of transverse heat current as a response to a longitudinal temperature gradient, which can be carried not only by electrons \cite{Y. Shiomi, W. Long, N. M. R. Peres}, but also by magnons \cite{Y. Onose, S. Murakami}, phonons \cite{T. Saito, C. Strohm, T. Qin} and photons \cite{C. R. Otey}.
Recently, the thermal Hall effect has been extended to nonlinear regime and an intrinsic nonlinear thermal Hall effect originated from Berry curvature is reported in clean systems with time-reversal symmetry \cite{C.-C. Zeng1}, in which a transverse heat current is generated vertically to the temperature gradient and scales quadratically with temperature gradient.
Usually, it is inevitable to hold defects and impurities in authentic materials. In this paper, we investigate an electron-carried nonlinear thermal Hall effect (NTHE) in presence of disorder by taking into account both intrinsic and extrinsic contributions in time-reversal invariant and noncentrosymmetric materials (fig.~\ref{fig1}).
In our study, the temperature dependence of the induced nonlinear thermal Hall conductivity (NTHC) $\kappa_{yxx}$ is found to be $\kappa_{yxx}^\text{in} \varpropto T^{2}$, $\kappa_{yxx}^\text{sj} \varpropto T^{0}$, $\kappa_{yxx}^\text{sk} \varpropto T^{0}$ at low temperature, which also provides a new approach to distinguish the intrinsic contribution ($\propto T^{2}$) from the extrinsic contribution ($\propto T^{0}$) that can be easily tested experimentally.

Within the framework of semi-classical theory, the total thermal Hall currents in absence of an electric field as response to temperature gradient in presence of a nontrivial Berry curvature $\Omega\left(\mathbf{k}\right)$ and disorder (see details in Ref.~\cite{Supple}) is found to be
\begin{equation}
\begin{aligned}
\mathbf{j}_{Q}^{T} &=\sum_{n}\int[d \mathbf{k}][E_{n}(\mathbf{k})-\mu]f_{l}(\mathbf{v}+\mathbf{v}_{l}^{s j})\\&+\frac{\nabla T}{T} \times \frac{1}{\hbar} \sum_{n}\int[d \mathbf{k}][E_{n}(\mathbf{k})-\mu] \mathbf{\Omega_{n}(\mathbf{k})}\\ & \times \left[(E_{n}(\mathbf{k})-\mu) f_{l}+k_{\mathrm{B}} T \log \left(1+e^{-\beta[E_{n}(\mathbf{k})-\mu]}\right)\right],
\label{zong}
\end{aligned}
\end{equation}
where $l=( n,\mathbf{k})$ is a combined index with the energy band $n$ and momentum $\mathbf{k}$,  $E_{n}(\mathbf{k})$ denotes the energy dispersion, $\mathbf{v}$ represents group velocity, $\mu$ indicates the chemical potential (Fermi energy), $\mathbf{v}_{l}^{s j}=\sum_{l^{\prime}} \varpi_{l l^{\prime}} \delta \mathbf{r}_{l^{\prime} l}$ is side-jump velocity originated from a disorder-induced coordinate shift $\delta \mathbf{r}_{l^{\prime} l}$, and the non-equilibrium distribution function $f_{l}$ can be expressed as $f_{l}=f_{l}^{in}+\delta f_{l}^{sk}$ in presence of temperature gradient and disorder, where $\delta f_{l}^{sk}$ is skew-scattering induced modification. The total heat current $\mathbf{j}^{T}_{Q}$ in Eq.~(\ref{zong}), thus, can be decomposed into four parts as $\mathbf{j}_{Q}^{T}=\mathbf{j}^{\text{N}}_{Q}+\mathbf{j}_{Q}^{in}+\mathbf{j}_{Q}^{sj}+\mathbf{j}_{Q}^{sk}$ corresponding  to the conventional, intrinsic, side-jump and skew-scattering contributions to heat current, respectively (details can be found in Ref.~\cite{Supple}). It has been found that the standard conventional heat current $\mathbf{j}^{\text{N}}_{Q}$ comes from the conventional velocity in time-reversal invariant materials~\cite{Supple}.

The linear thermal Hall current as a first-order response to temperature gradient will disappear in time-reversal invariant materials. That's because the linear thermal conductivity will be forced to be a symmetric tensor by the Onsager's reciprocity relations in the presence of time-reversal symmetry \cite{Landua1}.
The linear heat current will be aligned to the direction of temperature gradient, hinting the vanishing linear thermal Hall current. Therefore, only the nonlinear heat current possibly flows vertically to the temperature gradient in presence of time-reversal symmetry.
Through solving the Boltzmann equation, the non-equilibrium distribution function $f_{l}$ to the second-order correction to the temperature gradient can be determined \cite{Supple}. Accompanying with the formulas of $\mathbf{j}_{Q}^{in}$, $\mathbf{j}_{Q}^{sj}$ and $\mathbf{j}_{Q}^{sk}$ in Ref.~\cite{Supple}, the nonlinear thermal Hall current $\mathbf{j}^{\text{nl}}$ (where the superscript ``{nl}" refers to nonlinear) in the $a$ direction, as the response to the second order in temperature gradient, is found to be \cite{Supple}
\begin{equation}
j_{a}^{\text{nl}}\equiv -\kappa_{abd}\partial_{b}T\partial_{d} T,
\label{kappa}
\end{equation}
where $\kappa_{abd}$ is a coefficient quantizing the nonlinear thermal Hall effect and its formulas from the three mechanisms are given in Ref.~\cite{Supple}. Analogous to the nonlinear anomalous Hall effect \cite{Sodemann} and the nonlinear anomalous Nernst effect \cite{X.-Q. Yu} induced by Berry curvature dipole, it is found that only the states near the Fermi surface make contributions to the intrinsic NTHE, which is in contrast to the extrinsic contribution. Through the Sommerfeld expansion (see details in Ref.~\cite{Supple}), the coefficients $k_{abb}$ due to the intrinsic, side-jump and skew-scattering at low temperature, are found to be, respectively
\begin{eqnarray}
\kappa_{abb}^{\text{in}}&=&\left.-\frac{7 \tau\pi^{4}k_{B}^{4}}{15\hbar^{2}}T^{2}G_{0}^{\prime}(\mu)\right.+O[(k_{B}T)^{4}],\label{L-in} \\
\kappa_{abb}^{\text{sj}}&=&\left.-\frac{1}{3}\tau^{2}\pi^{2}k_{B}^{2}F_{0}(\mu)\right.+O[(k_{B}T)^{2}],\label{L-sj} \\
\kappa_{abb}^{\text{sk}}&=&\left.\frac{\tau^{3}\pi^{2}k_{B}^{2}}{3\hbar}g_{0}(\mu)\right.+O[(k_{B}T)^{2}]. \label{L-sk}
\end{eqnarray}
where $G_{0}\left(\mu \right)$, $F_{0}(\mu)$ and $g_{0}(\mu)$ (given in Ref.~\cite{Supple}) are Fermi energy dependent parameters and independent of temperature.
According to Eqs.~(\ref{L-in})-(\ref{L-sk}), one finds that the leading order of NTHC from these three contributions has following temperature $T$ dependence: $\kappa_{yxx}^{\text{in}} \varpropto T^{2}$, $\kappa_{yxx}^{\text{sj}} \varpropto T^{0}$, $\kappa_{yxx}^{\text{in}} \varpropto T^{0}$.
Through this temperature dependence, the intrinsic contribution ($\propto T^{2}$) can be easily distinguished from the extrinsic contribution ($\propto T^{0}$), which provides a new approach to identify the intrinsic and extrinsic mechanisms in experiments.
This new approach is totally different from the previous scaling law in which the dependence of anomalous Hall resistivity on the longitudinal resistivity are applied to distinguish the intrinsic and extrinsic contributions.

By exploiting the symmetry analysis, the nonvanishing NTHC can exist in time-reversal invariant and inversion symmetry broken materials. Hence, the non-centro-symmetric monolayer transition-metal dichalcogenides (TMDCs) is one of candidates to observe the nonlinear thermal Hall effect and the low-energy structure of TMDCs can be described by a simple tilted 2D massive Dirac model \cite{Du}, namely
\begin{equation}
\hat{\mathcal{H}}_{0}=t k_{x}+v\left(k_{x} \sigma_{x}+k_{y} \sigma_{y}\right)+m \sigma_{z},
\label{H0}
\end{equation}
where $\sigma_{x,y,z}$ are the Pauli matrices, and $t, v$ and $m$ are related model parameters. $t$ is a band titling parameter which tilts the Dirac cone along the $x$ direction, and $2m$ is the gap. The time reversal of the model contributes equally to the nonlinear thermal Hall current, so it is sufficient to study this model only. This Hamiltonian only considers one Dirac cone. Indeed, there is another inequivalent Dirac cone (the time reversal counterpart of this model) which, however, contributes equally to the nonlinear thermal Hall current. 
$\hat{\mathcal{H}}_{0}$ is invariant under the mirror operation $M_{x}$ about $x-y$ plane. The energy eigenvalues is $E_{n}(\mathbf{k})=t k_{x} +n E_{0}(\mathbf{k})$ with $E_{0}(\mathbf{k})=\sqrt{v^{2} k^{2}+m^{2}}$ where $n=\pm 1$ in $E_{n}(\mathbf{k})$ represents the band index. The Berry curvature is $\Omega_{\mathbf{k}}^{n}=-n \frac{m v^{2}}{2\left(E_{0}\left(\mathbf{k}\right)\right)^{3}}$.  In fact, in the $x-y$ plane, the Berry curvature is a pseudovector, and only the $z$ component exists $\Omega_{z}^{n}$.

\begin{figure}[tb]
\centering
\includegraphics[width=0.44\textwidth]{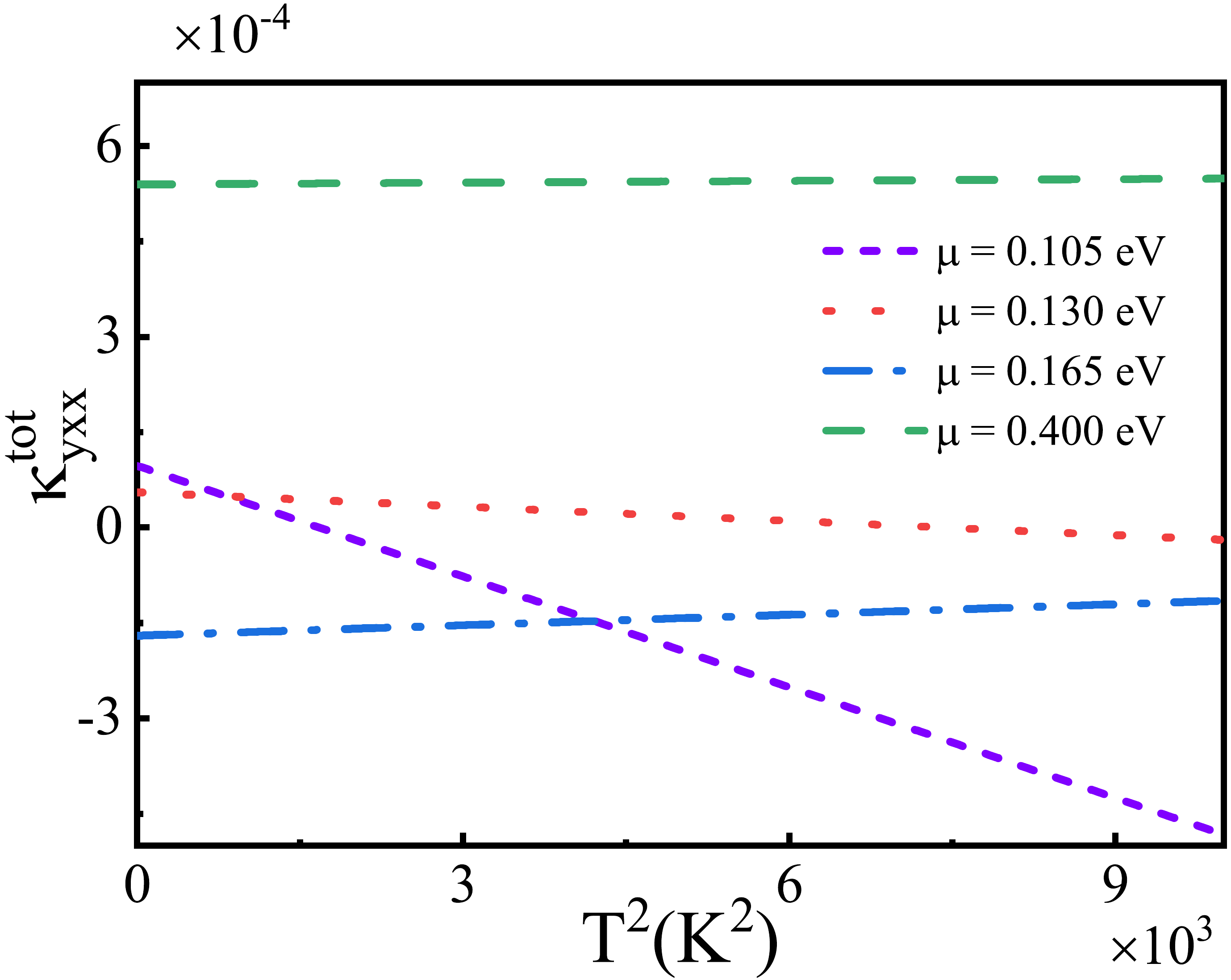}
\centering
\caption{Total nonlinear thermal Hall conductivity $\kappa_{xxy}^{tot}$ versus temperature at different Fermi energy. Here the unit for the $y$-axes is $k_{B}^{2}\cdot ${\AA}$/\hbar$. Parameters are taken as $t=0.1  ~\text{eV}\cdot ${\AA}$, \nu=1 ~\text{eV}\cdot ${\AA}$, m=0.1 ~\text{eV}, n_{i}V_{0}^{2}=10^{2} ~\text{eV}^{2}\cdot ${\AA}$^{2}$ and $n_{i}V_{1}^{3}=10^{4} ~\text{eV}^{3}\cdot ${\AA}$^{4}$.}
\label{fig2}
\end{figure}

In order to investigate the extrinsic contribution to the NTHE induced by disorder, we consider a $\delta$-function random potential $\hat{V}_{\text{imp}}(\mathbf{r})=\sum_{i} V_{i} \delta\left(\mathbf{r}-\mathbf{R}_{i}\right)$ with $\mathbf{R}_{i}$ indicating the random position of impurities and $V_{i}$ representing the disorder strength, which satisfies $\left\langle V_{i}\right\rangle_{\text{dis}}=0,\left\langle V_{i}^{2}\right\rangle_{\text{dis}}=V_{0}^{2} \neq 0$ and $\left\langle V_{i}^{3}\right\rangle_{\text{dis}}=V_{1}^{3} \neq 0$ \cite{N. A.}. To analyse the behaviour of NTHE of the tilted 2D massive Dirac model, we consider  $t\ll v$ limit and treat the relaxation time $\tau$ momentum independent, namely $\frac{1}{\tau}=\frac{n_{i} V_{0}^{2}}{4 \hbar} \frac{\mu^{2}+3 m^{2}}{v^{2} \mu}$ \cite{Du}, where $n_{i}$ represents the impurity concentration.

After a careful derivation (see details in Ref. \cite{Supple}), we find $\kappa_{xyy}^{\text{in}/\text{sk}/\text{sj}}= 0$ and $\kappa_{yxx}^{\text{in}/\text{sk}/\text{sj}}\neq0$. This suggests that only when applying temperature gradient in the $x$-direction perpendicular to the mirror line $M_{x}$ (the subscript $x$ means that the mirror plane is perpendicular to the x direction), there is a nonzero nonlinear thermal Hall current due to both the intrinsic and extrinsic mechanisms in the $y$ direction (perpendicular to the temperature gradient and parallel to the $M_{x}$).
In other words, once applying temperature gradient in the $y$-direction, the nonlinear thermal Hall current will disappear as required by the mirror symmetry M$_{x}$. The nonzero $\kappa_{yxx}$ to the first order of band tilting strength $t$ in intrinsic, side-jump and skew-scattering for the tilted 2D massive Dirac model in presence of disorder at low temperature are derived analytically as,
\begin{equation}
\begin{aligned}
\kappa_{yxx}^{\text{in}}&=\frac{7 \pi^{3}tmk_{B}^{2}}{5\hbar n_{i}V_{0}^{2}}\frac{v^{2}(\mu^{2}-2m^{2})}{\mu^{4}(\mu^{2}+3m^{2})}(k_{B}T)^{2}, \\ 
\kappa_{yxx}^{\text{sj}}&=-\frac{\pi tmk_{B}^{2}}{6\hbar n_{i}V_{0}^{2}}\frac{v^{2}(\mu^{2}-m^{2})(5\mu^{2}-9m^{2})}{\mu^{2}(\mu^{2}+3m^{2})}\label{k-in-k-sj},
\end{aligned}
\end{equation}
and
\begin{equation}
\begin{aligned}
\kappa_{yxx}^{\text{sk}}&=\frac{\pi v^{2} tm k_{B}^{2}}{3\hbar n_{i}^{2}V_{0}^{6}/V_{1}^{3}}\frac{(\mu^{2}-m^{2})(\mu^{2}-2m^{2})}{\mu(\mu^{2}+3m^{2})^3}\\&+\frac{\pi v^{2} tm k_{B}^{2}}{2\hbar n_{i}V_{0}^{2}}\frac{(\mu^{2}-m^{2})(3\mu^{2}-5m^{2})}{\mu^{2}(\mu^{2}+3m^{2})^3},
\end{aligned}
\label{k-sk}
\end{equation}
respectively. According to Eqs.~(\ref{k-in-k-sj}) and (\ref{k-sk}), at low temperature, the nonlinear thermal Hall coefficient from the intrinsic and extrinsic mechanisms can be identified through the temperature dependence of $\kappa^\text{tot}_{yxx}$  since the intrinsic $\kappa_{yxx}^\text{in}$ displays a quadratic dependence on temperature whereas both $\kappa_{yxx}^\text{sj}$ and $\kappa_{yxx}^\text{sk}$ are independent of temperature. The total $\kappa^\text{tot}_{yxx}$ can be written as $\kappa^\text{tot}_{yxx}=\alpha T^{2}+\lambda$ where the coefficients $\alpha$ and $\lambda$ represent the contribution from intrinsic and extrinsic mechanisms, respectively, and can be measured in future experiments.

\begin{figure*}[htbp]
\centering
\includegraphics[width=0.90\textwidth]{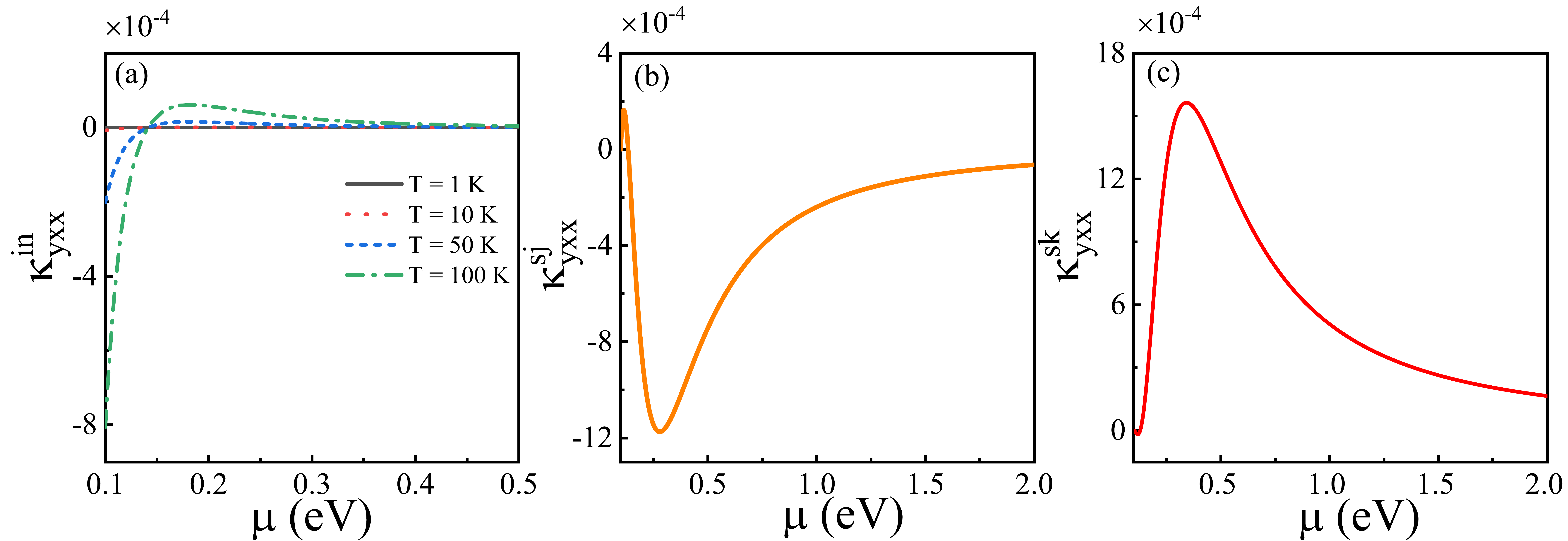}
\centering
\caption{Intrinsic (a), side-jump (b) and skew-scattering (c) contribution to the nonlinear thermal Hall conductivity $\kappa_{yxx}$ in tilted 2D massive Dirac model versus Fermi energy at different temperature, respectively. Here the unit for the $y$-axes is $k_{B}^{2}\cdot ${\AA}$/\hbar$, the other parameters are the same as in Fig. \ref{fig2}.}
\label{fig3}
\end{figure*}

Fig. \ref{fig2} displays the total NTHC $\kappa_{yxx}^\text{tot}$ as a function of $T^{2}$ at different Fermi energies for the titled 2D massive Dirac model. The intercept ($\lambda$) and the slope multiplied by the square of temperature ($\alpha T^{2}$) give the magnitudes of the extrinsic-mechanism-induced $\kappa^\text{ex}_{yxx}=\kappa^\text{sj}_{yxx}+\kappa^\text{sk}_{yxx}$ and intrinsic-mechanism-induced $\kappa^\text{in}_{yxx}$, respectively.
The enhanced intercept and decreased slope at higher Fermi energy indicate that the extrinsic contribution is gradually dominant. The signal of $\kappa_{yxx}$ from the extrinsic contribution is almost $245$ times larger than that from intrinsic contribution when the Fermi energy is taken at $0.4~$eV and the temperature is fixed at $50~$K.
However, when the Fermi energy is near the bottom of conduction band (e.g., $\mu=0.105~$eV), the slope becomes sharper and the intrinsic contribution is strengthened. It is observed that the magnitude of $\alpha T^{2}$ at $50~$K is equal to $1.49 \lambda$ for $\mu=0.105~$eV, implying the both contributions from intrinsic and extrinsic mechanisms to NTHE become comparable.

Besides the temperature dependence, the Fermi energy dependence of the NTHC can be used to identified these three scattering mechanisms (see Fig. \ref{fig3}). For $\mu=0.1$ eV, the Fermi level is just entering into the bottom of conduction band,
$\kappa_{yxx}^{\text{sj}}$ and $\kappa_{yxx}^{\text{sk}}$ tend to zero and only $\kappa_{yxx}^{\text{in}}$ is finite [Figs. \ref{fig3}(a)-(c)], revealing that all the heat current flowing vertically to the temperature gradient at Dirac point would stem solely from intrinsic contribution.
%
%
%
This band-edge-case may be a supplemental identification to the T-dependence identification. Since the Fermi level can be tuned in the TMDCs by a gate voltage, it is instructive to study the Fermi energy dependence further.  $\kappa_{yxx}^{\text{in}}$ drops dramatically with increasing $\mu>0.1$ eV and change its sign from negative to positive (about $\mu=0.14$ eV). It grows into a peak and finally decays to zero at large $\mu$.   $\kappa_{yxx}^{\text{sj}}$ is almost negative when Fermi level entering into the conduction band, and it develops a peak around $\mu=0.28$ eV and finally tends to zero.   The variation of $\kappa_{yxx}^{\text{sk}}$ on the Fermi energy is similar to that of $\kappa_{yxx}^{\text{sj}}$ but with an opposite sign. The peak of $\kappa_{yxx}^{\text{sk}}$ is around $\mu=0.34$ eV. Combining all the three components of $\kappa_{yxx}$, the total $\kappa_{yxx}^{\text{tot}}$ starts a negative value ($\mu=0.1$ eV, intrinsic component), drops dramatically with increasing $\mu$ and change to positive one, develops to a positive peak, and finally decays to zero at large $\mu$. It is seen that the positive $\kappa_{yxx}^{\text{tot}}$ is basically coming from the side-jump and skew-scattering contribution but the skew-scattering dominates.


In summary, we studied the nonlinear thermal Hall effect in presence of disorder as a second-order response of temperature gradient. 
Remarkably, it is revealed that, at low temperature, the intrinsic-mechanism-induced nonlinear thermal Hall conductivity is proportional to the square of temperature, while the extrinsic-mechanism-induced nonlinear thermal Hall conductivity is independent of temperature, which presents a new approach to clarify the intrinsic and extrinsic contributions. We also analysed the nonlinear thermal Hall effect for a tilted 2D Dirac material with mirror symmetry $M_{x}$. It is found that the mirror symmetry $M_{x}$ has a strong limitation on the nonlinear thermal Hall conductivity: only when the temperature gradient is applied perpendicular to the mirror line, a finite nonlinear thermal Hall current can be generated. Moreover, it is shown that as the Fermi energy entering into the bottom of the conduction band, the signal is solely from intrinsic mechanism. At high Fermi level, the signal from extrinsic mechanism (mainly skew-scattering) will dominate.


This work is supported by NSFC (Grant Nos. 11974348, 11674317, and 11834014) and the Fundamental Research Funds for the Central Universities. G.S. and Z.G.Z. are supported in part by the National Key R\&D Program of China (Grant No. 2018FYA0305800), the Strategic Priority Research Program of CAS (Grant Nos. XDB28000000, and XDB33000000). X.Q.Y is supported by NSFC (Grant Nos.12004107) and the Fundamental Research Funds for the Central Universities.


\end{document}